# Phase engineering of odd parity state in a quantum dot Josephson junction

Shohei Kobayashi[1,2], Tyler Lindemann[3], Sergei Gronin[3], Geoffrey Gardner[3],
Peter Stano[2,4], Daniel Loss[4,8,9,10,11], Michael Manfra[3,5,6,7], Seigo Tarucha[2,4], Sadashige Matsuo[1,2]

[1]*Department of Physics, Institute of Science Tokyo, Ookayama 152-8551, Meguro, Tokyo, Japan*
[2]*RIKEN Center for Emergent Matter Science, RIKEN, Hirosawa 2-1, Wako-shi, 351-0198, Saitama, Japan*
[3]*Birck Nanotechnology Center, Purdue University, West Lafayette, Indiana 47907, USA*
[4]*RIKEN Center for Quantum Computing, RIKEN, Hirosawa 2-1, Wako-shi, 351-0198, Saitama, Japan*
[5]*Department of Physics and Astronomy, Purdue University, West Lafayette, Indiana 47907, USA*
[6]*School of Materials Engineering, Purdue University, West Lafayette, Indiana 47907, USA*
[7]*School of Electrical and Computer Engineering, Purdue University, Indiana 47907, USA*
[8]*Department of Physics, University of Basel, Klingelbergstrasse 82, Basel CH-4056, Switzerland*
[9]*Physics Department, King Fahd University of Petroleum and Minerals, 31261, Dhahran, Saudi Arabia.*
[10]*Quantum Center, KFUPM, Dhahran, Saudi Arabia.*
[11]*RDIA Chair in Quantum Computing.*

kobayashi.s.36c5@m.isct.ac.jp
tarucha@riken.jp
matsuo@phys.sci.isct.ac.jp

**Controlling fermion parity in quantum-dot Josephson junctions (QDJJs) enables access to novel superconducting physics. While parity transitions are usually tuned electrically, we study phase-induced transitions in a strongly coupled QDJJ using the superconducting phase difference across the dot terminals, denoted by $\phi$. We find that this phase allows control over the ground state parity, with parity transitions observed as the Andreev bound states (ABSs) energies crossing zero. These crossings occur near $\phi = \pi$, and their behavior is in good correspondence to a simple model considering a single-level quantum dot proximitized by a superconductor. These results place phase-controlled QDJJ arrays as a promising platform for exploring exotic phenomena such as Majorana zero modes.**

## Introduction

Superconducting devices provide a versatile and tunable platform for exploring exotic superconducting (SC) phenomena such as the SC diode effect [1–4] and Majorana zero modes [5–8]. Among them, a quantum dot Josephson junction (QDJJ), a Josephson junction (JJ) having a single quantum dot (QD) between two SC leads, is a prototypical device structure, offering gate control as an additional handle in experiments [9–14]. Since electrical charge is not conserved in the SC systems, the states of such QDJJ are labelled not by fermionic occupation numbers, but by the fermion parity, which can be either even or odd. Control over the ground-state parity, which has been pursued experimentally [13–17], emerges as a basic requirement for the application of QDJJs to nontrivial tasks. To give two examples, the *SC diode effect* relies on a QDJJ with the odd-parity ground state [18–20], while a controllable one-dimensional array of QDJJs can emulate the *Kitaev chain*, offering access to topological superconductivity and Majorana zero modes [5,21–28]. Demonstrating controllable parity transition, and especially the odd-parity ground state, in a QDJJ is thus often a goal pursued in their experimental realizations.

Parity control in a QDJJ is typically approached through analysis of the Andreev bound states (ABSs) spectrum [12,16,29,30]. The ABS energies correspond to parity-changing excitation energies of the QDJJ. A transition in the system ground-state parity then corresponds to ABSs energies crossing zero. Since the system has particle-hole symmetry (inversion of energy about the Fermi energy), the ABSs appear in pairs reflected about the zero-energy axis. We conclude that a parity transition is detected as a symmetrically positioned pair of ABSs crossing zero energy upon changing system parameters.

Along these lines, parity transitions have been experimentally observed in quantum dots coupled to a single or a pair of SC leads [12,13,15–17,29]. In these experiments, the odd-parity ground state has been induced by gates, either tuning the QD energy level ($\varepsilon$) relative to the Fermi energy of the SC leads or the coupling strength ($\Gamma$) between the QD and the SC leads [12,13,15–17] and in Ref. [29], in addition, by exploiting the Zeeman energy. The latter was necessary as in the regime of strong coupling ($\Gamma > U$, with $U$ the dot charging energy), which applied to the experiment, a simplified model (the large superconducting-gap limit, $\Delta \to \infty$) shows that without the magnetic field the parity remains even [15,31,32].

In addition to these mechanisms, it has been theoretically predicted that the ground-state parity can be controlled by the phase difference of the SC condensate across the JJ [30–35], denoted here by $\phi$. Such phase-induced parity transitions have previously been reported in multiterminal Josephson junctions [36–38] and in a QDJJ embedded in a transmon circuit using microwave spectroscopy [14]. Complementary to these studies, we employ tunnel spectroscopy to directly probe phase-induced parity transitions in a QD strongly coupled to the SC leads and further investigate how Zeeman effect modifies the parity-transition boundaries. Utilizing tunnel spectroscopy, we examine the junction ABS energies as a function of the phase difference (Andreev spectra) [12,29,39–42] and observe that the lowest ABS pair crosses the zero energy line upon varying $\phi$ nearby the point $\phi = \pi$. We also examine the effects of the Zeeman splitting induced by an in-plane magnetic field and observe that it enhances the range of phases $\phi$ for which the system has the odd-parity ground state. Further supported by the good correspondence to a simple theoretical model, we conclude that an odd-parity ground state is realized in a QDJJ in the strong-coupling regime and have demonstrated control over the parity through gates, the superconducting phase, and the Zeeman energy.

## Device characteristics

We have fabricated JJ devices made from an InAs quantum well covered with a thin epitaxial Al film [43,44]. The fabricated device structure is shown in Fig. 1(a). The JJ was defined by selective wet etching of Al, resulting in a 60-nm distance between two SC leads, and embedded in a SC loop to control the phase difference by an out-of-plane magnetic field $B_{\perp}$. Some of the measurements reported below use also an in-plane magnetic field $B_{||}$ , whose direction is indicated by the arrow in Fig. 1(a). To perform tunneling spectroscopy and form the QD within the SC leads, three gate electrodes were patterned on the quantum well. Two gate electrodes, t1 and t2 (with voltages $V_{\mathrm{t1}}$ and $V_{\mathrm{t2}}$ that are both fixed at -3.22 V in all measurements below) are adjacent to the JJ and form a gate-tunable tunnel barrier between the QD and the 2-dimensional electron gas (2DEG). The third gate (with voltage $V_{\mathrm{g}}$) pinches off the 2DEG between the two SC electrodes, except for an approximately 70-nm-wide area highlighted by a red dashed line. With this setup, a QD well coupled to the SC leads can be formed in the red-highlighted region (60 nm × 70 nm) and can be biased with respect to 2DEG by applying a voltage $V_{\mathrm{bias}}$ between the InAs and the superconductor.

We measure the differential conductance $G$ through the tunnel barrier using the standard lock-in technique, with all measurements done in a dilution refrigerator at a base temperature of 50 mK. In the following paragraphs, we explain the following procedure: We first establish the quantum dot, its coupling to the superconductor, access to its Andreev spectrum, and the calibration of the phase difference $\phi$, fixing the point $\phi = 0$. With these arrangements in place, we analyze the ground-state parity in detail, examining its response to various system parameters.

## Fundamental device properties

As the first step, we suppress superconductivity by applying an out-of-plane field (500 mT) exceeding the critical field of the Al film (~10 mT). The Al leads become normal, and we measure $G$ as a function of $V_{\mathrm{bias}}$ and $V_{\mathrm{g}}$, plotting it in Fig. 1(b). We observe conductance peaks, which we interpret as the Coulomb peaks of a QD tunnel coupled to effectively two leads with highly asymmetric couplings: one lead is formed by the InAs 2DEG and the other by the Al film. Observing Coulomb diamonds confirms that a QD has been formed. The QD is weakly coupled to the InAs 2DEG but strongly coupled to the Al film. The strong coupling broadens the QD levels, while the asymmetric couplings lead to unequal visibility of the Coulomb-diamond edges. These effects result in smeared and apparently distorted Coulomb diamonds and are consistent with the QDJJ operating in the strong-coupling regime.

We next turn on the superconductivity by repeating the measurement at a small out-of-plane magnetic field ($B_{\perp}$ below 1 mT). We tune the QD to a point well inside a Coulomb diamond, using an appropriate value of $V_{\mathrm{g}}$ identified in the normal state measurement, and scan the magnetic field and the SC phase. For these magnetic fields, the Al remains superconducting and the only effect of $B_{\perp}$ is to induce a phase difference $\phi$ across the JJ, by threading a magnetic flux through the loop formed by the Al film.

Figure 1(c) shows the resulting $G$. We observe two pronounced conductance resonances symmetrically with respect to $V_{\text{bias}} = 0$ mV. The slight offset of the spectral center from zero bias voltage is a measurement artefact, while the visible asymmetry between the currents at positive and negative energies is due to energy-dependent tunneling matrix elements. We observe the SC gap, and conductance resonances outside of it, oscillating periodically with $B_{\perp}$. The oscillation period estimated from the data is 0.16 mT, which agrees well with the period of $\Phi_0/S \approx 0.172$ mT expected based on the SC loop area $S \approx 12\ \mu\text{m}^2$. Since the system energies are periodic in phase $\phi$ with period $2\pi$, the observed period (we use 0.16 mT) allows us to convert a change in the magnetic field $B_{\perp}$ into a change in the phase $\phi$. To fully determine the conversion, we also need to calibrate the point $\phi = 0\ (\text{mod}\ 2\pi)$. (From here on, we omit the "(mod $2\pi$)" when stating values for phase). It can differ from $B_{\perp} = 0$ mT applied externally because of a small remnant magnetic field or flux trapped in the superconductor. We exploit the fact, confirmed by the model below, that SC gap observed in a figure such as Fig. 1(c) is maximal for $\phi = 0$. Figure 1(d) shows a close-up, from which one can see that indeed, $B_{\perp} = 0$ is much closer to $\phi = \pi$, than to $\phi = 0$, and thus the unknown offset is not negligible. In any case, the offset can be quantified from the figure, and we can thus reliably convert the experimentally controlled $B_{\perp}$ parameter to the physically relevant parameter $\phi$.

Apart from the calibration, the data scans, such as the one plotted in Fig. 1(d), serve as a probe of the QDJJ parity and its transitions. Namely, the lowest ABS pair, positioned symmetrically with respect to the horizontal axis, $V_{\text{bias}} = 0$, reveals the lowest energy cost of a ground-state excitation. The spectrum plotted in Fig. 1(d) as a function of $\phi$ displays the expected phase dependence of the ABSs in QDJJ [12,29,32,40,45], thereby confirming that the device is ready to probe the parity transition in a phase-controlled QDJJ.

## Parity transition in Andreev spectrum measurement

Figure 2(a) shows such an examination scan, after changing the voltage from the value $V_g$ compared to from the value used for Fig. 1(c). Upon this change, we now observe two crossings of ABSs, at $\phi = \phi_{c1}$ and $\phi_{c2}$, indicated by two white circles. Although not obvious at first glance, the ABSs look inverted (compare with Fig. 3(b) below) in the region between the two crossings $\phi_{c1} < \phi < \phi_{c2}$. This reversal, stemming from the two consecutive crossings of the ABSs energies, is an experimental signature of the parity transition [12,13,17,29]. We conclude that inside the region $\phi_{c1} < \phi < \phi_{c2}$, the QDJJ has the different parity from that outside the region.

To display the phase-induced parity transition more clearly, as well as to further investigate it, in Fig. 2(b) we show the conductance measured at zero bias as a function of $\phi$ and $V_g$. The bright curves in Fig. 2(b) correspond to phase values $\phi$ at which the positive and negative energy ABSs cross and thus delineate the boundaries between different parities. We observe the parity transition boundary taking the shape of a loop centered at $\phi = \pi$ and $V_g = -2.195$ V. We do not observe any inversions nearby $\phi = 0$, which is consistent with strong coupling between the QD and the SC leads, even across wider sweeps of the gate voltage (see supplementary information). We thus conclude that in our strongly coupled QDJJ, the parity transition occurs within a limited $\phi$ range around $\phi = \pi$.

To understand these experimental results in more detail, we model the QDJJ in the large superconducting-gap limit using an effective Hamiltonian $H$ [32]:

$$H = \sum_{\sigma} \epsilon d_{\sigma}^{\dagger} d_{\sigma} + U d_{\uparrow}^{\dagger} d_{\uparrow} d_{\downarrow}^{\dagger} d_{\downarrow} - \Gamma \cos\left(\frac{\phi}{2}\right) \left(d_{\uparrow}^{\dagger} d_{\downarrow}^{\dagger} + d_{\downarrow} d_{\uparrow}\right) \quad (1)$$

Here, $d_{\sigma}^{\dagger}$ and $d_{\sigma}$ are, respectively, creation and annihilation operators of an electron in the QD with spin direction denoted $\sigma = \uparrow, \downarrow$. The eigenstates of this Hamiltonian can be obtained analytically as follows:

odd-parity state $|\uparrow\rangle$, with energy $E_{\uparrow} = \epsilon$, (2a)

odd-parity state $|\downarrow\rangle$, with energy $E_{\downarrow} = \epsilon$, (2b)

even-parity state $|+\rangle = u|\downarrow\uparrow\rangle + v|0\rangle$, with energy $E_{+} = \left(\epsilon + \frac{U}{2}\right) + E_d$, (2c) even-parity state $|-\rangle = -v|\downarrow\uparrow\rangle + u|0\rangle$, with energy $E_{-} = \left(\epsilon + \frac{U}{2}\right) - E_d$. (2d)

In these equations, we have used quantities defined by $E_d^2 = \left(\epsilon + \frac{U}{2}\right)^2 + \Gamma^2 \cos^2(\phi/2)$, $u^2 = [E_d + (\epsilon + \frac{U}{2})]/2E_d$ , and $v^2 = [E_d - (\epsilon + \frac{U}{2})]/2E_d$. The ground-state parity is even or odd, depending on which of the four eigenstates has the lowest energy, and will be a function of the parameters $\epsilon, U, \Gamma$, and $\phi$. Assuming strong coupling and finite charging energy ($\Gamma > U > 0, -U < \epsilon < 0$), the eigenenergies are plotted as a function of $\phi$ in Fig. 3(a). We observe that $E_{-}$ crosses the degenerate pair $E_{\uparrow}, E_{\downarrow}$ at two points ($\phi = \phi_{c1}, \phi_{c2}$) around $\phi = \pi$. The parity of the model system is odd ($|\uparrow\rangle, |\downarrow\rangle$ have the lowest energy) in $\phi_{c1} < \phi < \phi_{c2}$, and even ($|+\rangle$, $|-\rangle$ has the lowest energy) outside of this range.

To enable direct comparison with experiment, we note that tunnel spectroscopy probes excitations involving a single fermion transfer, that is, a transition between many-body states of different parity. The presence of such an excitation at an energy equal to $eV_{\mathrm{bias}}$ results in an increased tunneling current. Accordingly, we re-express the eigenenergies of the many-body system described by Eq. (1) in terms of excitation energies between different parity states. Since the states $|\uparrow\rangle$ and $|\downarrow\rangle$ are degenerate, there are two such parity-changing excitation energies. To mimic the particle-hole symmetry of the system, we reflect the two energies about the horizontal axis, yielding four excitation energies shown in Fig. 3(b). The ABS spectrum obtained in this way closely resembles the experimental ABS spectra shown in Fig. 2(a) and supports our conclusion that we indeed experimentally achieved the phase-induced parity transition in a strongly coupled QDJJ and that, within the region $\phi_{c1} < \phi < \phi_{c2}$ in Fig. 2(a), the system has odd parity.

Having established the connection to the experiment, we use the model in Eq. (1) to further investigate the expected parity as a function of the system parameters. In Fig. 3(c), we plot the system "parity phase diagram" as a function of $\phi$ and $\Gamma$. We observe that the system has odd parity as long as the ratio $\Gamma/U$ remains below a threshold that depends on the phase $\phi$. The threshold is smallest for $\phi = 0$, where it equals $1/2$, grows as the phase is displaced from $\phi = 0$, and diverges as the phase approaches

$\phi = \pi$.

In Fig. 3(d), we inspect the phase diagram further, this time as a function of parameters $\phi$, and $\epsilon$. Importantly, the loop structure similar to the one experimentally obtained in Fig. 2(b) is reproduced in this diagram and assigns odd parity to the interior of the loop.

## Zeeman effect in parity transition

Finally, we explore spin-related effects on the parity transition. To this end, we apply an in-plane magnetic field $B_{\parallel}$ and repeat the measurements that yield Fig. 2(b) discussed above. The results are shown in Fig. 4. First, a measurement analogous to that reported in Fig. 2(a) is repeated at a slightly different gate voltage, with the results shown in Fig. 4(a) serving as a $B_{\parallel} = 0$ mT benchmark. The odd-parity region is highlighted by the red arrow. The gate condition in Fig. 4 differs from that in Fig. 2 because the measurement was performed after a deep negative gate sweep, which shifted the effective gate-voltage offset owing to gate hysteresis and a charge jump. Nevertheless, the zero-bias conductance map confirmed that an odd-parity region was realized under the gate condition used for Fig. 4. We then increase the in-plane magnetic field and observe how the transition points change. In Fig. 4(b) we show one particular spectrum, at a fixed magnetic field of $B_{\parallel} = 400$ mT. The odd-parity region becomes larger. The increase in the extent of the odd-parity region is clearly visible in Fig. 4(c), where we plot the zero-bias conductance as a function of $\phi$ and $B_{\parallel}$. The white dashed lines placed at parity transition points suggest that the odd-parity region linearly expands as $B_{\parallel}$ increases.

This behavior can be easily understood from the simple model in Eq. (1), once it is supplemented with a Zeeman term, with $E_z$ being the Zeeman energy:

$$H' = H + \sum_{\sigma} \frac{E_z}{2} \sigma\, d_{\sigma}^{\dagger} d_{\sigma} \tag{3}$$

The eigenstates of the Hamiltonian remain the same as before, and the only consequence of the new term is that it splits the energies of the two odd-parity states $E_{\uparrow,\downarrow} = \epsilon \pm \frac{E_z}{2}$. The degeneracy splitting is visible in Fig. 4(d), where we replot the Hamiltonian spectrum previously shown in Fig. 3(a). The figure shows that, irrespective of the other parameters, such a splitting enlarges the odd-parity region whenever it exists. This is demonstrated in Fig. 4(e), which is a plot analogous to Fig. 3(b). In the even-parity regions, $\phi < \phi_{c1}$ and $\phi > \phi_{c2}$, the green curves in Fig. 4(e), $|E_- - E_\uparrow|$, represent possible parity-changing excitations from the ground state. However, these branches are not clearly resolved experimentally, possibly because of the limited energy resolution of the tunneling spectroscopy, magnetic-field-induced spectral broadening, and small transition matrix elements.

Finally, Fig. 4(f) shows the model prediction of the parity transition boundary using the same plot coordinates as in Fig. 4(c). The Zeeman effect lowers the energy $E_\downarrow$ of the odd-parity state linearly with $E_z$, causing $\phi_{c1}$ to shift to lower phase and $\phi_{c2}$ to shift to higher phase. Consequently, the phase interval $\phi_{c1} < \phi < \phi_{c2}$, corresponding to the odd-parity ground state, expands with $E_z$, reproducing the approximately linear growth observed over the investigated range.

## Conclusion

In summary, we have investigated the Andreev spectra of QDJJs in the regime of strong coupling between the QD and the leads. For specific values of the gate voltages, we have observed inverted ordering of the ABSs' energies close to the point where the superconducting phase difference across the dot is $\pi$. Based on a comparison with a simple model (a single-level proximitized quantum dot), we have attributed this inversion to parity transitions in the system ground state. We have also observed that applying an in-plane magnetic field increases the phase interval over which the system is in the odd-parity ground state. Overall, these results demonstrate that an odd-parity state in a QDJJ can be realized via phase control even in the strong-coupling regime. This finding makes QDJJ arrays in the strong-coupling regime a promising platform for engineering exotic phenomena, such as topological SC and the SC diode effect, with phase control.

## Figures

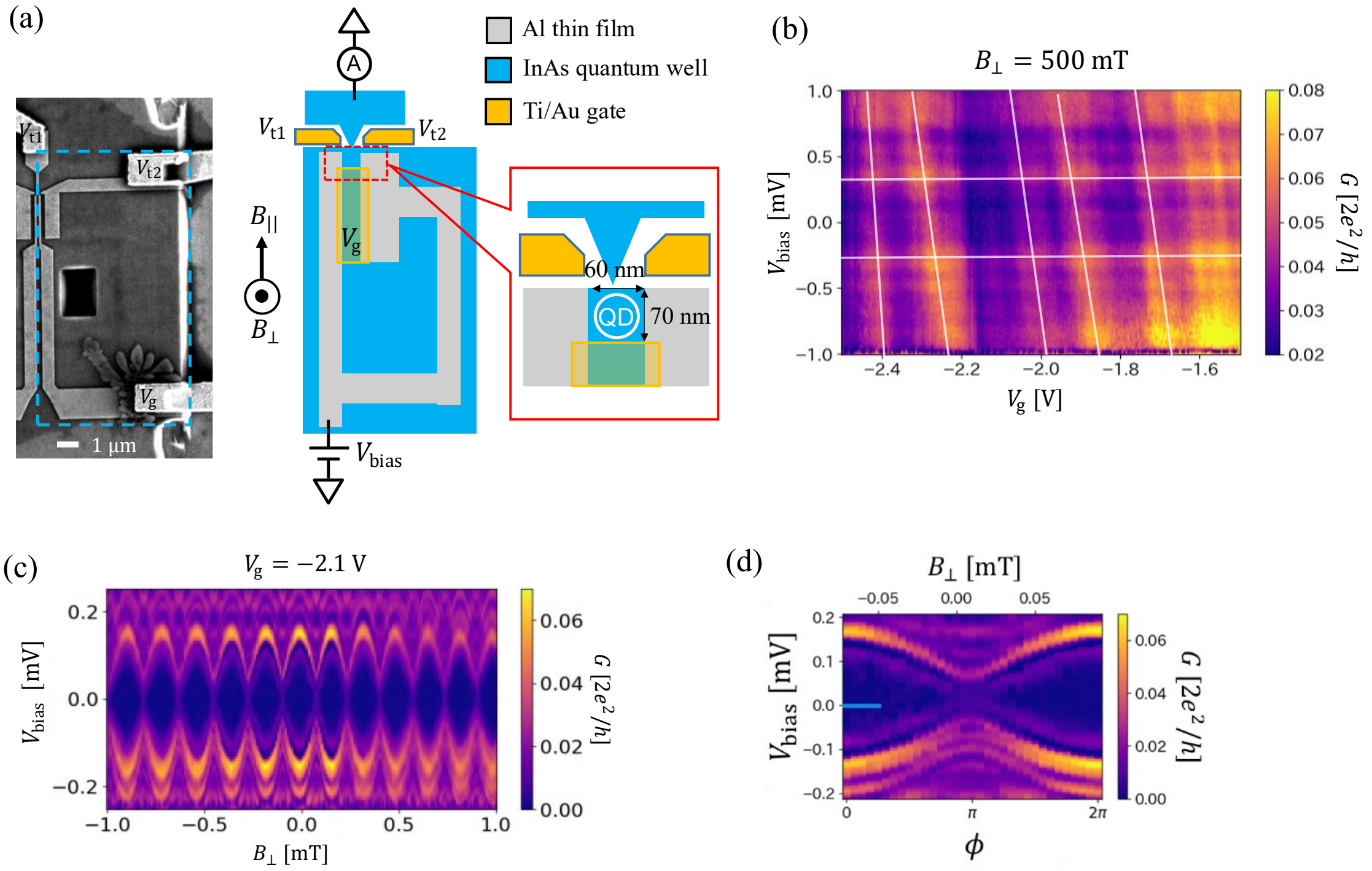

**Fig. 1. Device structure and tunnel spectroscopy measurements**

(a) SEM micrograph of the fabricated device, together with a schematic illustration of the device structure and measurement setup. The SEM image includes a 1 μm scale bar. The differential conductance through the tunnel barrier formed by $V_{t1}$ and $V_{t2}$ is measured. Electrons in the bare InAs quantum well between the two SC electrodes are pinched off by $V_g$ except for an approximately 70-nm-wide area highlighted by the red dashed line. A QD is thereby formed electrically between two SC leads.

(b) $G$ as a function of $V_{bias}$ and $V_g$ at $B_\perp = 500$ mT, which is larger than the critical field of Al thin film. The Coulomb peaks, highlighted by the white dashed lines, are observed, indicating the formation of a QD.

(c) $G$ as a function of $B_\perp$ and $V_{bias}$ for $-1.0\ \text{mT} \leq B_\perp \leq 1.0\ \text{mT}$ at $V_g = -2.1$ V. Two pronounced conductance resonances appear symmetrically with respect to $V_{bias} = 0$ V. These resonances correspond to ABSs whose energies depend on the phase difference between the two SC leads.

(d) $G$ as a function of $B_\perp$ and $V_{bias}$ for $-0.07\ \text{mT} \leq B_\perp \leq 0.09\ \text{mT}$. $B_\perp$ (the upper x axis) is converted to the phase difference $\phi$ (the lower x axis).

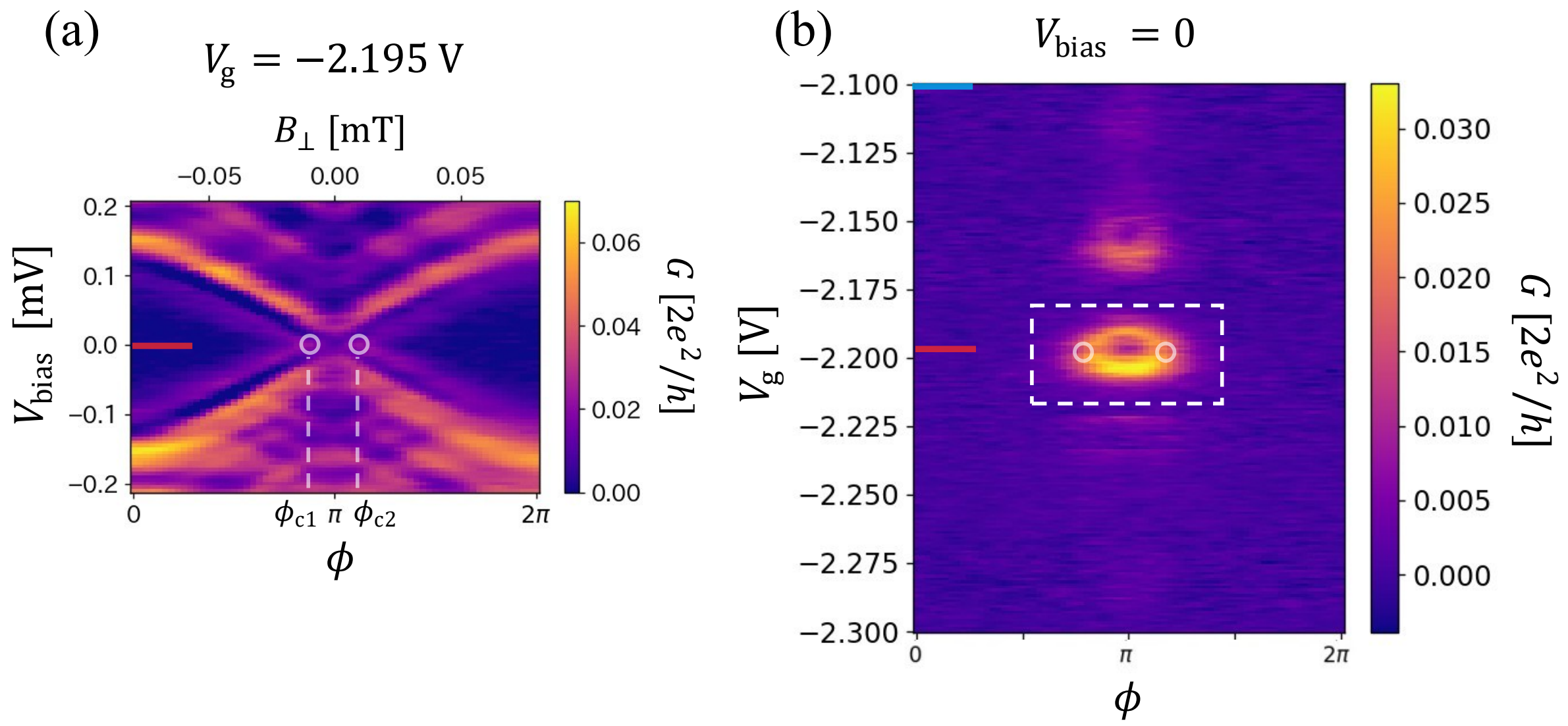

**Fig. 2. Phase-induced parity transition**

(a) $G$ as a function of $B_\perp$ and $V_{bias}$ at $V_g = -2.195$ V. The positive and negative energy ABSs cross at two points ($\phi = \phi_{c1}, \phi_{c2}$) at $V_{bias} = 0$, highlighted by two white circles.

(b) $G$ as a function of $\phi$ and $V_g$ at $V_{bias} = 0$. The bright curves correspond to $\phi$ at which the positive and negative energy ABSs cross at zero bias. A loop structure around $V_g = -2.195$ V delineates the parity-transition points. The system parity inside the loop is thus different from the one outside the loop.

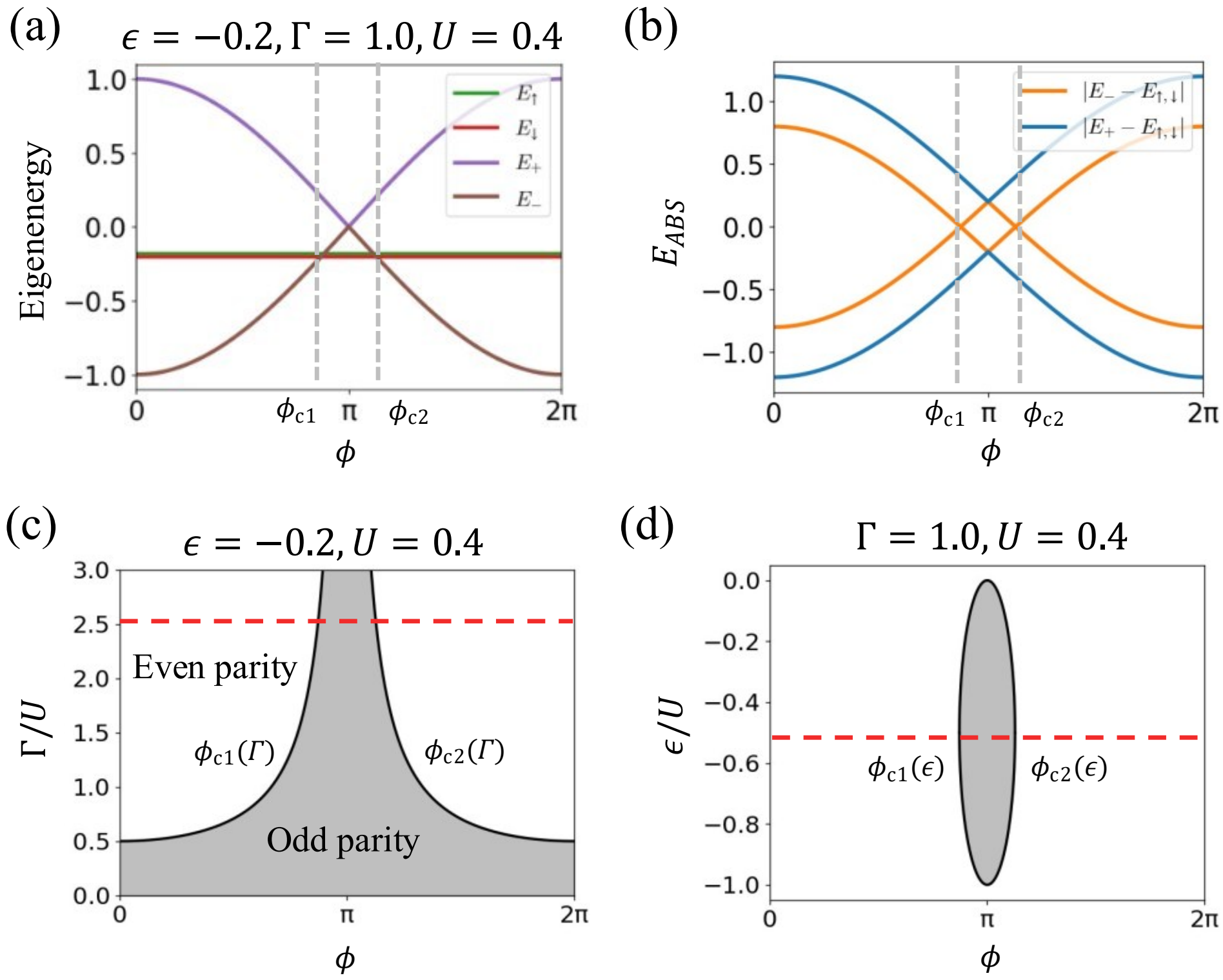

**Fig. 3. Calculated spectra and phase diagrams of the QDJJ model.**

(a) Eigenenergy ($E_\uparrow, E_\downarrow, E_+$, and $E_-$) calculated from the effective Hamiltonian in Eq. (1) as a function of $\phi$ for the four eigenstates, $|\uparrow\rangle$, $|\downarrow\rangle$ (odd-parity states), and $|+\rangle$, $|-\rangle$ (even-parity states). We used parameters $\epsilon = -0.2$, $\Gamma = 1.0$ and $U = 0.4$.

(b) The first and second excitation energies (the ABS energies) between the even-parity ($|+\rangle$, $|-\rangle$) and odd-parity ($|\uparrow\rangle$, $|\downarrow\rangle$) states as a function of $\phi$ calculated from eigenenergies in Fig. 3(a). The ordering of the lowest ABSs (orange solid lines) is inverted for $\phi_{c1} < \phi < \phi_{c2}$, indicating the odd parity within this range.

(c) A phase diagram of the parity state in the QDJJ as a function of $\phi$ and coupling strength $\Gamma$. The gray and white regions in the diagrams correspond to the odd- and even-parity states, respectively.

(d) A phase diagram of the parity states in the QDJJ as a function of $\phi$ and the QD energy level $\epsilon$.

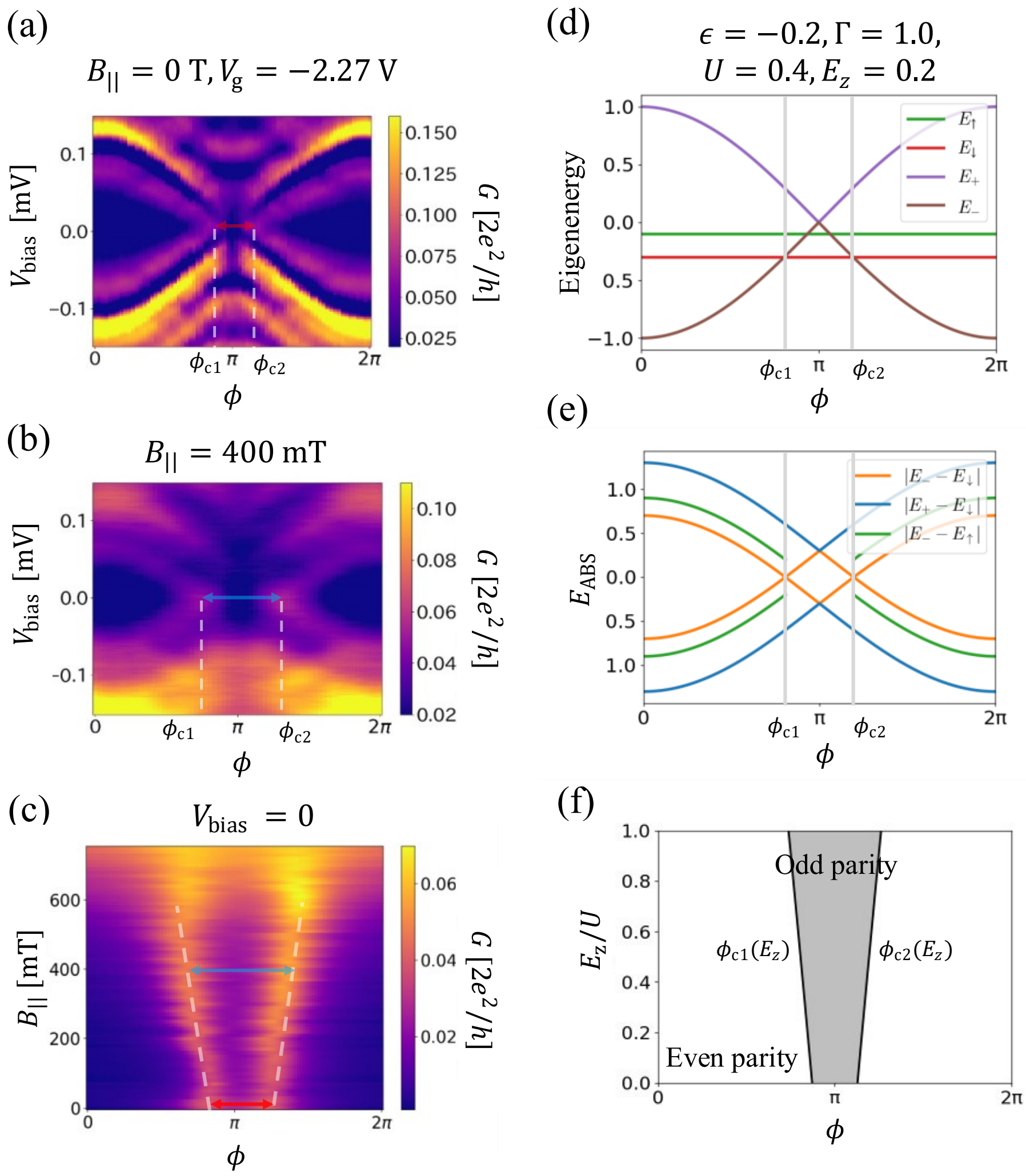

**Fig. 4. Phase-induced parity transition under an in-plane magnetic field.**

(a) $G$ as a function of $\phi$ and $V_{\text{bias}}$ at $B_{||} = 0$ mT with $V_g = -2.27$ V over one period. A parity transition is highlighted by the red arrow.

(b) $G$ as a function of $\phi$ and $V_{\text{bias}}$ at $B_{||} = 400$ mT over one period. The parity transition is highlighted by the blue arrow.

(c) $G$ as a function of $\phi$ and $B_{||}$ at $V_{\text{bias}} = 0$ V, $V_g = -2.27$ V. The dashed lines are guides to the eye, and approximately trace the conductance maximum, corresponding to the parity transition points.

(d) The eigenenergies of the model Hamiltonian $H'$ including the Zeeman energy. We used parameters $\epsilon = -0.2,\ \Gamma = 1.0,\ U = 0.4$, and $E_z = 0.2$.

(e) The excitation energies for possible parity-changing transitions from the ground state.

(f) Phase diagram of the parity as a function of $\phi$ and Zeeman energy $E_z$. The parity is calculated from spectra such as the one shown in Fig. 4(d). The odd parity region expands as $E_z$ increases, consistent with the experimental observation in Fig. 4(c).

# APPENDIX A: METHOD

**Sample growth**

The wafer has been grown on a semi-insulating InP substrate by molecular beam epitaxy. The stack materials from the bottom to top are a 100 nm $In_{0.52}Al_{0.48}As$ buffer, a 5 period 2.5 nm $In_{0.53}Ga_{0.47}As$/2.5 nm $In_{0.52}Al_{0.48}As$ superlattice, a 1µm thick metamorphic graded buffer stepped from $In_{0.52}Al_{0.48}As$ to $In_{0.84}Al_{0.16}As$, a 33 nm graded $In_{0.84}Al_{0.16}As$ to $In_{0.81}Al_{0.19}As$ layer, a 25 nm $In_{0.81}Al_{0.19}As$ layer, a 4 nm $In_{0.81}Ga_{0.19}As$ lower barrier, a 5 nm InAs quantum well, a 10 nm $In_{0.81}Ga_{0.19}As$ top barrier, two monolayers of GaAs and finally an 8.7 nm layer of epitaxial Al. The top Al layer has been grown in the same chamber without breaking the vacuum. The two-dimensional electron gas (2DEG) is accumulated in the InAs quantum well layer.

**Device Fabrication**

In this study, conventional electron beam lithography was used to fabricate JJs. We etched out the aluminum film using Type-D etchant after we defined the mesa of the InAs quantum well with 1:1:8 of $H_3PO_4$:$H_2O_2$:$H_2O$ etchant. Subsequently, we grew a 30 nm-thick $Al_2O_3$ film by atomic layer deposition and deposited Ti and Au to make the gate electrodes.

**Measurement setup**

For tunnel spectroscopy, we used a lock-in amplifier with an excitation frequency of 17 Hz. A DC bias voltage from a voltage source and an AC excitation voltage from the lock-in amplifier were summed and then attenuated by a factor of 1/1000 before being applied to the device. The tunnel current through the tunnel barrier was measured by the lock-in amplifier.

**Supplementary Note1: Device structure**

We refer to the device discussed in the main text as device A. Note that the schematic image in Fig.1(a) is simplified for readability. A schematic illustration and a scanning electron microscope image of the device without simplification are shown in Figs. 5(a) and (b). The two JJs share a common central superconducting (SC) electrode. The two JJs are respectively embedded in SC loops that enclose the same area. Three gate electrodes were placed adjacent to the two JJs to form two quantum point contacts, QPCL and QPCR, for tunnel spectroscopy. By tuning the gate voltages $V_{\mathrm{tR}}$, $V_{\mathrm{tC}}$, and $V_{\mathrm{tL}}$, we can selectively form either QPCL or QPCR and perform tunnel spectroscopy on either JJL or JJR. In the measurements presented in the main text, we performed tunnel spectroscopy on JJR through QPCR while pinching off QPCL and JJL by applying gate voltages $V_{\mathrm{tR}} = -5.0$ V, $V_{\mathrm{tC}} = V_{\mathrm{tL}} = -3.22$ V and $V_{\mathrm{gL}} = -4.0$ V. Therefore, we omit the QPCL, JJL, and the SC loop including JJL from Fig. 1(a). Instead of $V_{\mathrm{tC}}$, $V_{\mathrm{tR}}$, and $V_{\mathrm{gR}}$ in this note, we use $V_{\mathrm{t1}}$, $V_{\mathrm{t2}}$, and $V_{\mathrm{g}}$ in the main text.

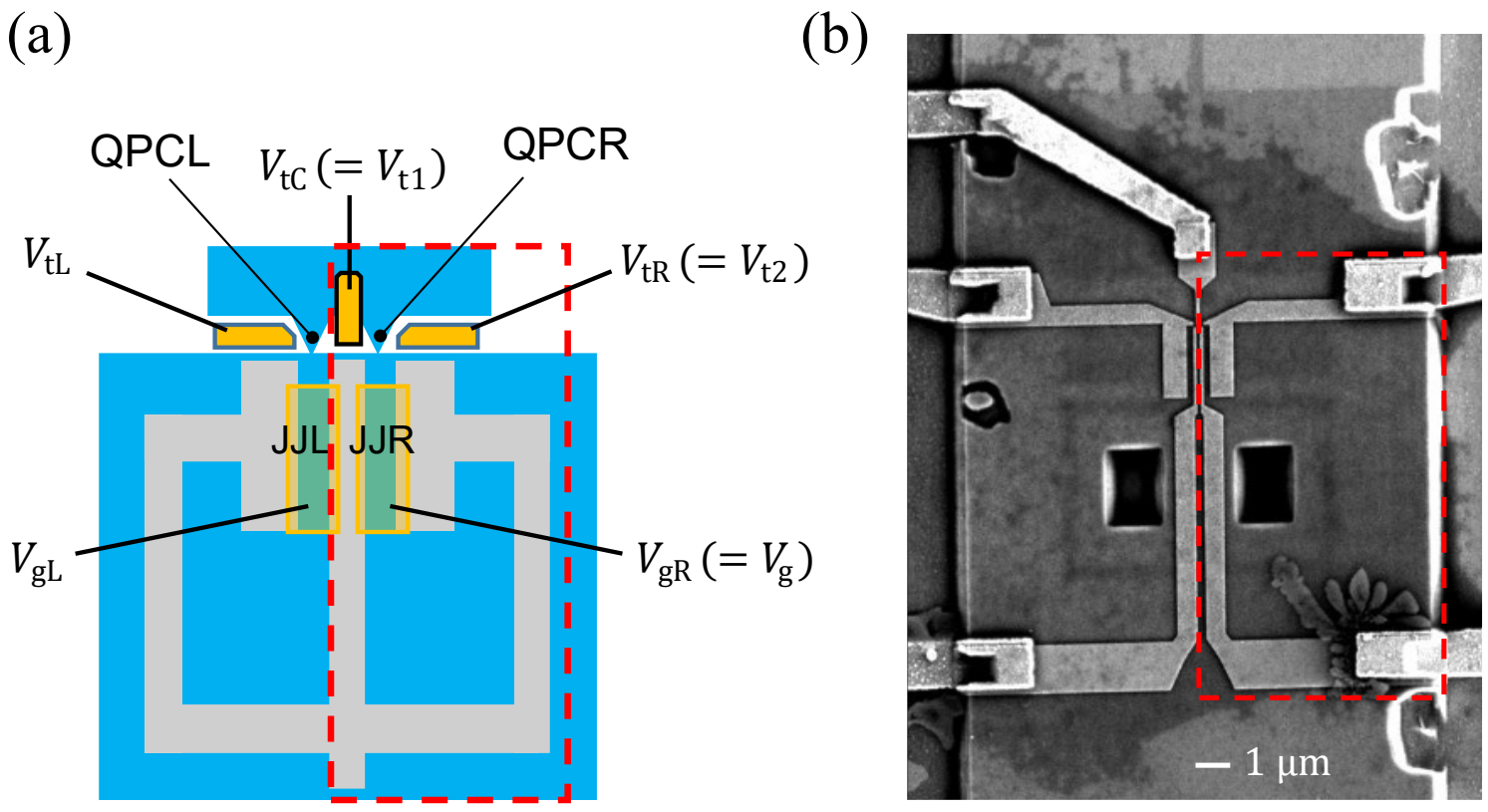

Fig. 5. Device structure
(a) A schematic image (b) A scanning electron microscopy image

## APPENDIX B: GATE VOLTAGE DEPENDENCE OF ABS AT $\phi$ = 0 AND $\pi$ IN DEVICE A

To confirm that the QDJJ is in the strong-coupling regime, we measured the gate-voltage dependence of ABSs at $\phi = 0$ and $\pi$ in the range of $-2.5\ \text{V} \leq V_g \leq -2.0\ \text{V}$, as shown in Figs. 6(a) and (b). This $V_g$ range corresponds to the region where Coulomb peaks appear in Fig. 1(b), indicating formation of a QD. As seen in Fig. 6(a), a crossing between the positive- and negative-energy ABS branches is not observed at $\phi = 0$, which indicates that the QDJJ is in the strong-coupling regime. In contrast, at $\phi = \pi$, crossings of the positive- and negative-energy ABS branches are observed at several $V_g$ values, as highlighted by the white arrows in Fig. 6(b). This is consistent with the phase-induced parity transition occurring around $\phi = \pi$ in the strong-coupling regime, as discussed in the main text.

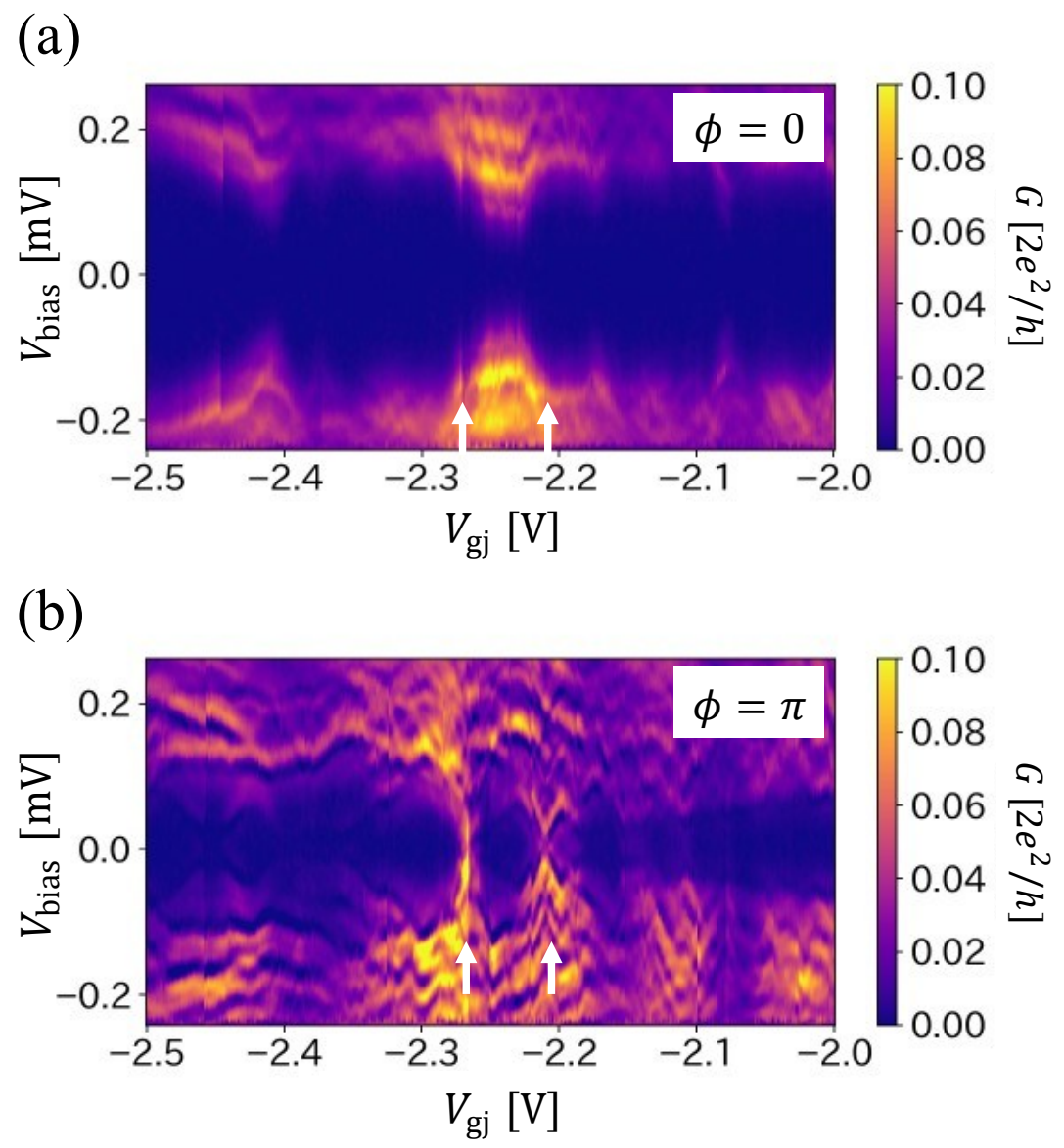

Fig. 6. Gate-voltage dependence of ABS
(a) $G$ as a function of $V_g$ and $V_{bias}$ at $\phi = 0$. The positive- and negative-energy ABS branches do not cross in this $V_g$ range, indicating that the QDJJ is in the strong-coupling regime.
(b) $G$ as a function of $V_g$ and $V_{bias}$ at $\phi = \pi$. The positive- and negative-energy ABS branches cross at several $V_g$ values as highlighted by the white arrows, indicating that a parity transition occurs by phase control in the strong-coupling regime.

## APPENDIX C: REPRODUCIBILITY IN A DIFFERENT DEVICE

We checked the reproducibility of the results using a different device (device B). To confirm reproducibility, we measured device B, which has almost the same structure as device A, except for an unintentionally broken SC loop. The structure of device B is depicted in Fig. 7. No supercurrent flows through JJL, and JJL does not affect the properties of JJR. We performed tunnel spectroscopy on JJR through QPCR while pinching off QPCL with $V_{\mathrm{tL}} = V_{\mathrm{tC}} = -4.6$ V, $V_{\mathrm{gL}} = 0.0$ V, and $V_{\mathrm{tR}} = -3.16$ V. Instead of $V_{\mathrm{tC}}$, $V_{\mathrm{tR}}$, and $V_{\mathrm{gR}}$ in this note, we use $V_{\mathrm{t1}}$, $V_{\mathrm{t2}}$, and $V_{\mathrm{g}}$ from here on.

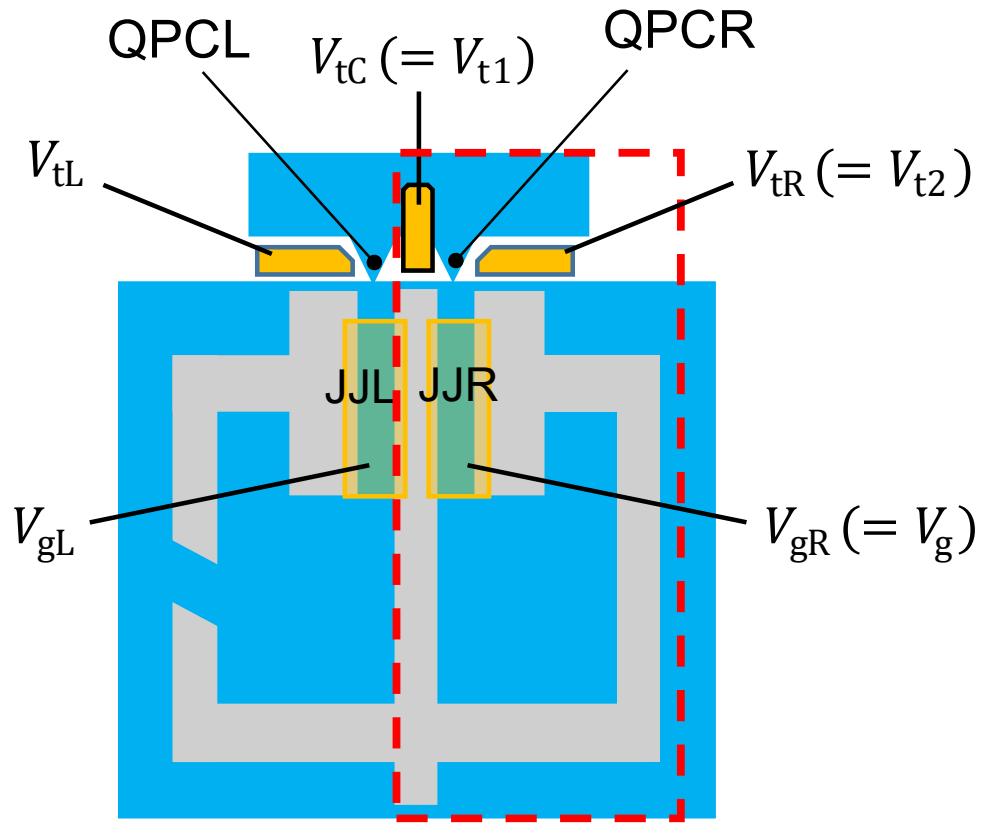

Fig. 7. Schematic image of device B

First, we examined the phase dependence of ABSs. Fig. 8(a) shows $G$ as a function of $V_{\mathrm{bias}}$ at $V_{\mathrm{g}} = -2.5$ V in the range of $-0.775\ \mathrm{mT} \leq B_{\perp} \leq 0.845\ \mathrm{mT}$. Two pronounced conductance resonances appear symmetrically with respect to $V_{\mathrm{bias}} = 0$ mV and oscillate periodically with $B_{\perp}$. Using the same procedure as in the main text, we convert $B_{\perp}$ to the phase difference $\phi$. The obtained spectrum as a function of $\phi$ clearly indicates expected phase dependence of the ABSs in QDJJ.

Next, we investigated the phase-induced parity transition. Fig. 8(b) shows $G$ as a function of $\phi$ and $V_{\mathrm{bias}}$ at $V_{\mathrm{g}} = -2.6$ V. As seen in device A, the positive- and negative-energy ABS branches cross at two points ($\phi = \phi_{\mathrm{c1}}, \phi_{\mathrm{c2}}$) at $V_{\mathrm{bias}} = 0$ indicated by two white circles in Fig. 8(b) and inverted to $V_{\mathrm{bias}} = 0$ V in $\phi_{\mathrm{c1}} < \phi < \phi_{\mathrm{c2}}$ where $\phi$ is around $\pi$. This clearly reproduces the result of the phase-controlled parity transition in the QDJJ shown in Fig. 2(a) of the main text.

We then fixed $V_{\mathrm{bias}} = 0$ V and measured $G$ as a function of $\phi$ and $V_{\mathrm{g}}$ as shown in Fig. 8(c). The bright curves in Fig. 8(c) correspond to $\phi$ at which the positive and negative ABSs cross, and thus indicate the boundaries between the different parity states. Sweeping $V_{\mathrm{g}}$ does not induce any inversion of the ABSs at $\phi = 0$, which is consistent with strong coupling between the QD and the SC leads. A loop structure appears around $V_{\mathrm{g}} = -2.6$ V around $\phi = \pi$, reproducing the results shown in Fig. 2(b).

Therefore, we confirm the reproducibility of the phase-induced parity transition in the strong-coupling regime from the phase-dependent ABS spectra measured in a different device.

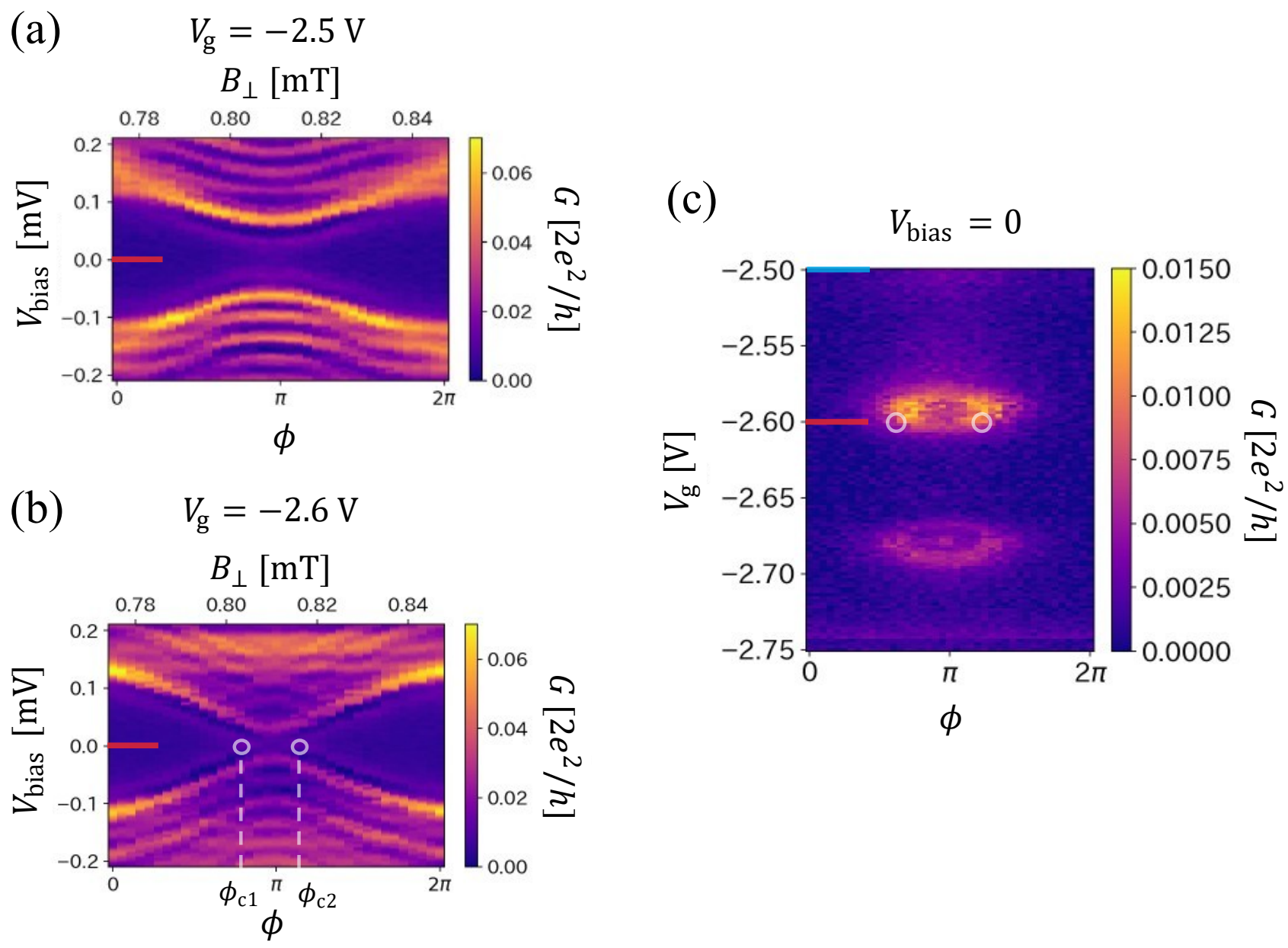

Fig. 8. Reproducibility of the experimental results in device B

(a) $G$ as a function of $B_\perp$ and $V_{bias}$ for $-0.775\ \text{mT} \leq B_\perp \leq 0.845\ \text{mT}$ at $V_g = -2.5$ V. Two pronounced conductance resonances appear symmetrically with respect to $V_{bias} = 0$ V. These resonances correspond to ABSs whose energies depend on the phase difference between the two SC leads. $B_\perp$ is converted to the phase difference $\phi$.

(b) $G$ as a function of $B_\perp$ and $V_{bias}$ for $-0.775\ \text{mT} \leq B_\perp \leq 0.845\ \text{mT}$ at $V_g = -2.6$ V. The positive- and negative-energy ABS branches cross at two points ($\phi = \phi_{c1}, \phi_{c2}$) at $V_{bias} = 0$ highlighted by two white circles and inverted in $\phi_{c1} < \phi < \phi_{c2}$, indicating the parity transition.

(c) $G$ as a function of $\phi$ and $V_g$ at $V_{bias} = 0$. The bright curves correspond to $\phi$ at which the positive and negative energy ABSs cross at zero bias. A loop structure around $V_g = -2.6$ V indicates that the parity inside the loop is different from that outside the loop.

## Acknowledgments

This work was partially supported by JST FOREST (Grant No. JPMJFR223A), the Murata Science Foundation, the CANON foundation, and the Asahi Glass foundation.